\documentclass[aps,prl,showpacs,twocolumn]{revtex4}
\usepackage{graphicx}
\begin{document}
\newcommand {\be}{\begin{equation}}
\newcommand {\ee}{\end{equation}}
\newcommand {\bea}{\begin{eqnarray}}
\newcommand {\eea}{\end{eqnarray}}
\newcommand {\nn}{\nonumber}

\title{Geometry fluctuations and Casimir effect in a quantum antiferromagnet}
\author{ Anuradha Jagannathan$^a$ and Attila Szallas$^b$ }
\affiliation{$^a$Laboratoire de Physique des Solides, CNRS-UMR 8502, Universit\'e
Paris-Sud, 91405 Orsay, France }
\affiliation{$^b$ Wigner Research Centre for Physics, Hungarian Academy of Sciences,
H-1525 Budapest, P.O.Box 49,  Hungary}
\date{\today}

\begin{abstract}
We show the presence of a Casimir type force between domain walls in a two dimensional Heisenberg antiferromagnet subject to geometrical fluctuations. The type of fluctuations that we consider, called phason flips, are well known in quasicrystals, but less so in periodic structures. As the classical ground state energy of the antiferromagnet is unaffected by this type of fluctuation, energy changes are purely of quantum origin. We calculate the effective interaction between two parallel domain walls, defining a slab of thickness $d$, in such an antiferromagnet within linear spin wave theory. The interaction is anisotropic, and for a particular orientation of the slab we find that it decays as $1/d$, thus, more slowly than the electromagnetic Casimir effect in the same geometry.

\end{abstract}
\maketitle

\begin{figure}[t]
\begin{center}
\includegraphics[scale=0.5]{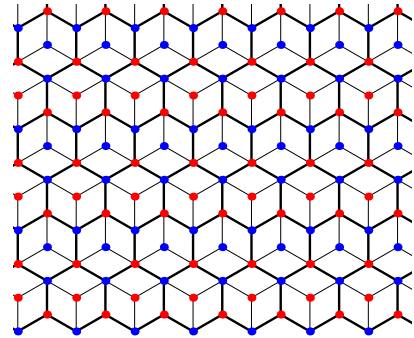}
\caption{ The staggered dice lattice. The sites are colored red or blue according to whether they belong in the A or B sublattice.}
\label{stdice.fig}
\end{center}
\end{figure}

The Casimir effect \cite{casimir} refers to the net force between two electrically neutral objects  that arises due to vacuum fluctuations of the electromagnetic field when they are placed suffiently close together. Initially considered as a somewhat mysterious apparition of ``something out of nothing'', the phenomenon described by Casimir is far more widespread, and arises quite generally due to fluctuations in media where long ranged correlations \cite{kardar2} are present. Thermal fluctuations in superfluids \cite{kardar1} and superconductors \cite{williams} give rise to long ranged forces. Casimir forces have been shown to exist in nematic liquid crystals \cite{ajdari}, or even granular systems \cite{cattuto}. In this paper we consider the interaction energy in a thin slab of the spin-$\frac{1}{2}$ antiferromagnetic Heisenberg model on the staggered dice lattice illustrated in Fig.\ref{stdice.fig}. The interfaces correspond to a row of phason flips of the structure, and hence to boundary conditions of a novel type compared to the usual Dirichlet or Neumann boundary conditions.

In our model, periodic boundary conditions are assumed so that there are no external boundaries. The parallel slab geometry in this antiferromagnet is instead determined by geometrical defects termed phason flips, in which the bonds of a single spin flip between two equivalent local configurations. We assume that the transitions between the two configurations can occur easily and that the magnitude of the nearest neighbor Heisenberg spin-spin couplings, J, is unaffected by the change. Such a situation could arise in structures in which there are two equivalent chemical pathways to generate antiferromagnetic Heisenberg superexchange, as we will discuss at the end of this paper. This type of disorder is reminiscent of the ''two level systems`` that were proposed for amorphous systems \cite{tls}, in order to explain the anomalously high scattering rate of phonons seen at low temperatures.
While phason flip disorder is well-known in the field of quasicrystals \cite{note}, its effects have only recently been considered in a periodic structure, namely the staggered dice antiferromagnet \cite{aj2012}. Compared to the disordered quantum spin models studied in the literature such as bond-disordered models or site-dilution models ~\cite{kato,sand,lin,mucc,flor}), phason flip disorder leads to some novel properties. In contrast to bond-fluctuation or site-dilution, a phason flip does not modify the classical ground state energy of the antiferromagnet. It does however modify the quantum corrections to the ground state energy and the spectrum of excitations around the ground state. As a result, antiferromagnetism was found to be $enhanced$ rather than $diminished$ as in other types of disordered antiferromagnets \cite{aj2012}. This phenomenon can be considered as a variant of the order-by-disorder effect originally found by Villain for a frustrated \cite{footnote} spin model ~\cite{villain}.

The Heisenberg model considered is
\bea
H=J \sum_{\langle i,j \rangle} \vec{ S}_i \cdot \vec{ S}_j
\label{heis.eq}
\eea
with $J>0$, where $\vec{S}_i$ are the spin operators for the sites $i$ of the staggered dice lattice and the sum runs over all pairs of spins $i,j$ joined by an
edge of length $a$. On the dice lattice, one can distinguish hexagonal ``cages" formed by
six spins, of which three are coupled to the spin at the cage center. This can occur in one of two ways.  The staggered dice lattice is formed by assembling such hexagonal cages such that the bonds alternate between the ''up" and ''down" configurations. The lattice vectors of the SDL shown in Fig.\ref{stdice.fig} are given by $\vec{r}_1=\sqrt{3}a\hat{x}$ and $\vec{r}_2=3a\hat{y}$, and the symmetry of the structure is rectangular.
The SDL structure is unfrustrated, as can be seen in Fig.\ref{stdice.fig}, where spins are colored red or blue, according to the sub-lattice and one expects rotational symmetry to be broken in the ground state, which is N\'eel ordered, with spins on sublattice A aligned along, say, the $+z$-direction, and spins on sublattice B aligned along $-z$. We recall that in the ordinary dice (also known as T3) lattice, the central sites are linked to three neighbors in a uniform way instead of this alternating staggered fashion, and that the result is a ferrimagnet \cite{prb2007}. In spin wave theory, one represents the spin operators in terms of Holstein-Primakoff boson operators $a_i$($a^\dag_i$) for sites $i$ on the A-sublattice and $b_i$($b^\dag_i$) for sites $i$ on the B-sublattice \cite{hols}. The linear spin wave Hamiltonian for the SDL lattice of N sites obtained from Eq.\ref{heis.eq} after neglecting higher order terms is

\bea
H_{LSW} &=& -2JS(S+1)N + H^{(2)} \nonumber \\
H^{(2)} &=& JS\sum_{\langle i,j \rangle} (a_i^\dag a_i + b_j b_j^\dag +a_i^\dag b_j^\dag +  a_ib_j)
\eea
One can introduce the Fourier transformed set of operators
\bea
a_{\mu k} = \frac{1}{\sqrt{N_c}}\sum_n e^{-i \vec{k}.\vec{R}_n} a_{\mu n} \\
b_{\mu k} = \frac{1}{\sqrt{N_c}}\sum_n e^{i \vec{k}.\vec{R}_n} b_{\mu n}
\eea
where the sum runs over the $N_c =N/6$ unit cells situated at positions $ \vec{R}_n$ and $\mu=1,2,3$ labels the three sites belonging to each sublattice within a unit cell.
The Hamiltonian now reads
\bea
H^{(2)} = JS\sum_k\sum_{\mu,\mu'} z_\mu \delta_{\mu\mu'}(a_{\mu k}^\dag a_{\mu k} + b_{\mu k} b_{\mu k}^\dag) + \nonumber \\
(\gamma_{\mu\mu'}(k) a_{\mu k}b_{\mu'k} + h.c.)
\eea
where $z_\mu=4,5$ and 3 are the coordination numbers for $\mu =1,2,3$ and $\gamma(k)$ is the $3\times 3$ matrix
\bea
\left(\begin{array}{cccccc}
 f_2^*(1+f_1^*) &1 & f_1^*f_2^* \\
1 & 1+f_1 & 1+f_1^* \\
f_2^* &1+f_1 &  0
\end{array} \right)
\eea
with $f_\mu = \exp(i \vec{k}.\vec{r}_\mu)$. We consider henceforth the solutions obtained for $S=\frac{1}{2}$, the case for which the quantum fluctuations are strongest. The energies of the spin excitations in the model are determined upon diagonalization by means of a Bogoliubov transformation. In the limit that $\vec{k}$ tends to zero one finds the linear relationship for the lowest energy Goldstone modes $\hbar\omega(\vec{k})= v_sk$ (for $k \ll a^{-1}$) where $v_s = 6\sqrt{\frac{2}{41}}J$ is the spin wave velocity. The ground state energy per site was calculated in linear spin wave theory to be
$E_0=-0.6517(7)J$ in the thermodynamic limit, which compares well to the result  $E_0=-0.6639(1)J$
obtained from quantum Monte Carlo (QMC) simulations \cite{aj2012}.

\begin{figure}[t]
\begin{center}
\includegraphics[scale=0.6]{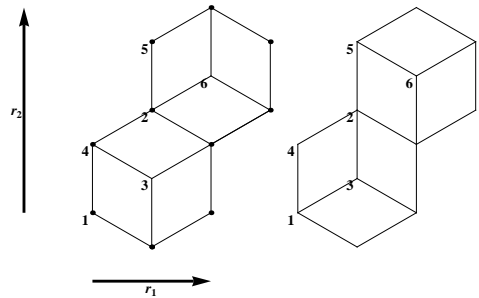}
\caption{Bond configurations before (left)  and after (right) two phason flips. The six sites of the unit cell are numbered.}
\label{flip.fig}
\end{center}
\end{figure}

Fig.\ref{flip.fig} illustrates the effect of a pair of phason flips within a single unit of the dice lattice. The interior spins $\vec{S}_3$ and $\vec{S}_6$ are coupled with three cage spins in different ways in the left hand and the right-hand figures. As the total number of bonds is constant, the total classical energy is unchanged by phason flips. As can be seen from Fig.\ref{flip.fig}, the transition between the two configurations implies a change of the sublattice of the center spin. In order to preserve the antiferromagnetic condition on the total spin, $S_{tot}=0$, the number of flips on sublattices A and B respectively, are therefore taken to be equal in all the (finite) systems that we consider. Since phason flips do not introduce any frustration, N\'eel antiferromagnetic order is always preserved and linear spin wave theory can thus be used to study the system within the entire range of values of the number of flips, $0\leq N_{f}\leq N/3$ where $N$ is the total number of spins. Periodic boundary conditions are assumed along $x$ and $y$ directions.

The energy cost per phason flip can be calculated in spin wave theory by a numerical diagonalization procedure. It was found to be approximately $0.06 J$, far smaller than, for example, the energy of a spin vacancy which is about $0.6J$ \cite{bulut}. The energy of interaction as a function of the distance has a short distance component -- of the order of one unit cell spacing --  and an anisotropic power law decay at long distance. The sign of the interaction energy of two phasons depends on the sublattice: for phasons on opposite sublattices, the interaction is attractive, and for those on the same sublattice it is repulsive. The rapid decay of perturbations around a phason flip can be understood in terms of a continuum version of the discrete Hamiltonian of Eq.\ref{heis.eq} by the fact that a flip produces gradient terms of high (third and above) order.

\begin{figure}[t]
\begin{center}
\includegraphics[scale=0.35]{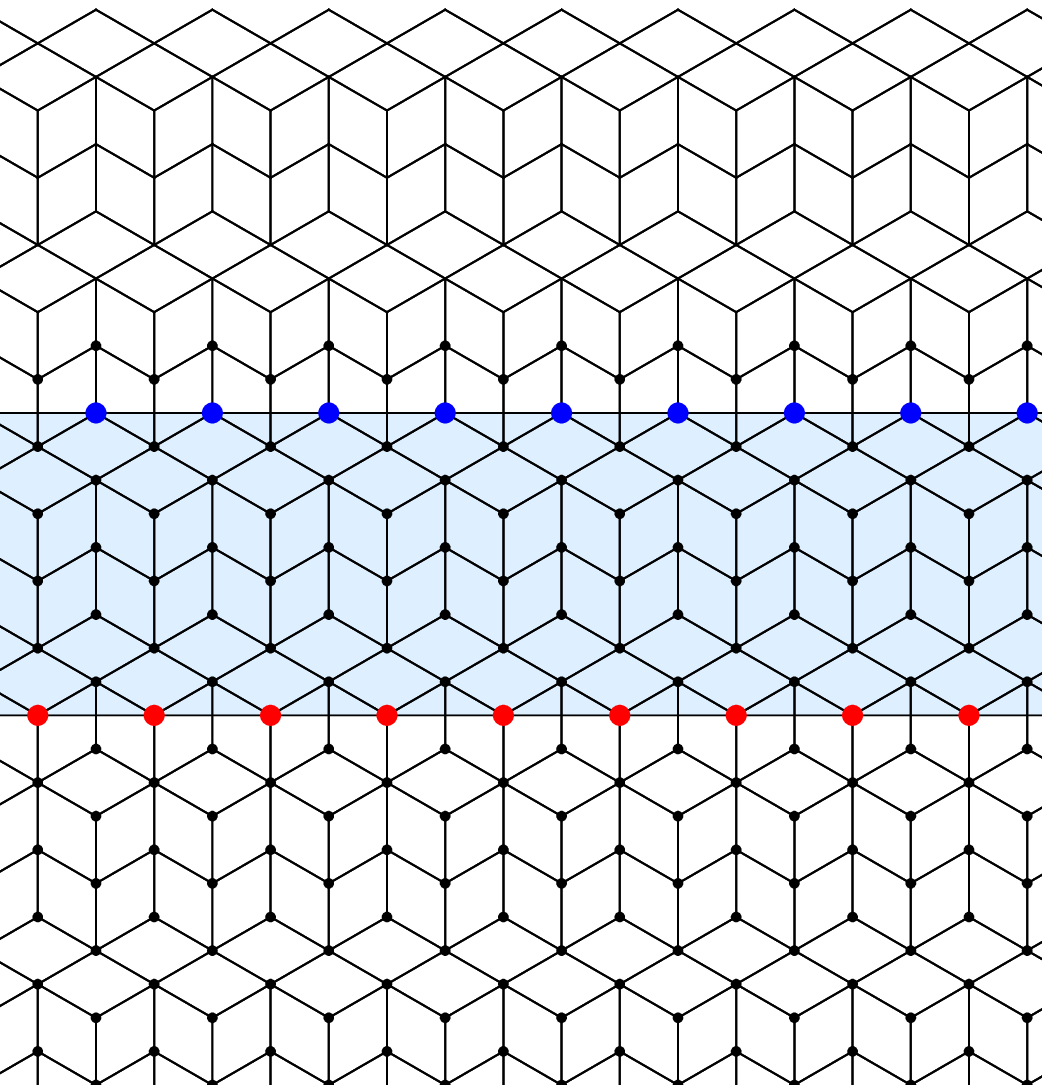}
\includegraphics[scale=0.35]{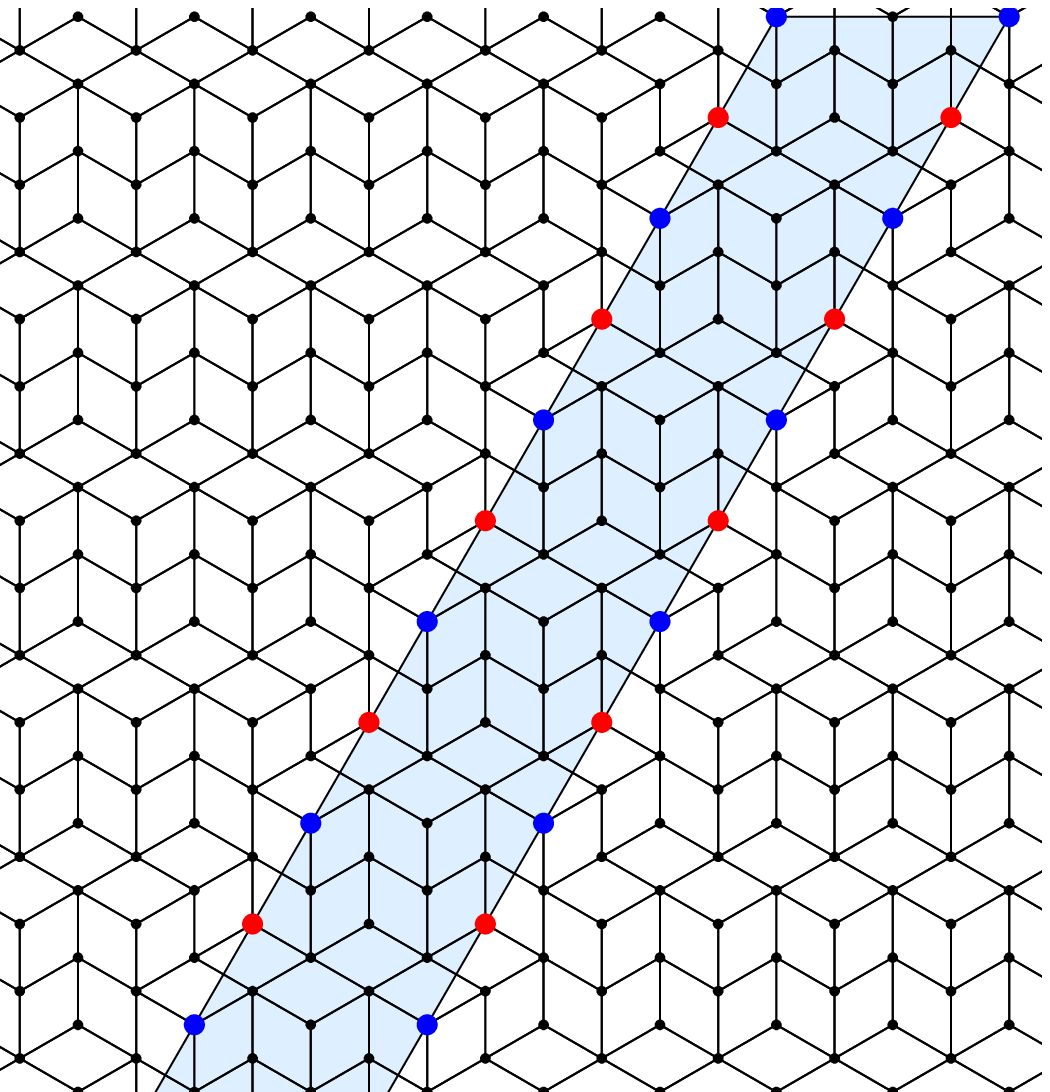}
\caption{
Illustration of domain wall configurations used in the study of Casimir effect. a) Slab aligned along $\vec{r}_1$ direction (shaded light blue) b) Slab aligned along $\vec{r}_1+\vec{r}_2$ direction.  Sites on the interfaces belong in either the A-sublattice (red points) or B-sublattice (blue points). }
\label{twolines.fig}
\end{center}
\end{figure}

We now present results for the energy dependence as a function of thickness for a horizontal (i.e. having its orientation parallel to the $\vec{r}_1$ direction) slab of thickness $d$. Fig.\ref{twolines.fig}a shows the configuration studied: the lower row of A-sublattice phasons is shown in red, while the upper row corresponds to B-sublattice phasons, shown in blue. The resulting system is analogous to a parallel plate capacitor: {{outside`` the parallel lines one has the original dice lattice, while ``inside'' the plates one has the fully flipped version of the original lattice. For each value of $d$, we calculated the difference of ground state energy $E_N(d)-E_0$ where $E_0$ is the ground state energy of the defect-free SDL. Dividing this by the length of the domain wall $L_x$, we obtain the energy per unit length, $\epsilon$. As Fig.\ref{casim.fig} shows, the interaction between the lines is attractive.
We fitted the energy change to the expression

\bea
\frac{(E_N(d)-E_0)}{L_x}= e_N(d) = A_N + B_N\left( \frac{1}{d}+\frac{1}{L_y-d}\right)
\label{casim.eq}
\eea
where the second term of the term in brackets arises due to the periodic boundary conditions ($L_y$ is the periodic length in the y-direction). Fitting to the data using Eq.\ref{casim.eq} for distances that are sufficiently large, we obtain the coefficients $A_N$ and $B_N$ for each sample size. The energy versus distance plot for three different sample sizes are shown in Fig.\ref{casim.fig}, along with the curves given by $A_N + B_N/d$ for $N=4800,5888$ and $6912$.  Both A and B coefficients follow a finite size scaling in $1/N$, in accordance with the scaling expected for the energy per length of a 1D antiferromagnet. Extrapolation to the infinite system yields the interface energy per length: $e(d) = A +B/d$, with the values $A=0.062 J/a$ and $B=0.001 J$. If one takes the exchange coupling $J$ to have a value comparable to that of typical Heisenberg antiferromagnets, namely a fraction of an $eV$, and that the edge length is of the order of an Angstrom, one obtains an interaction energy of about $1 pJ/m$.

When the horizontal slab calculations are repeated but for domain walls of phason flips on the $same$ sublattice, the resulting force is repulsive, with the same $d$ dependence. This behavior is consistent with the earlier studies of the interaction between two phason flips: repulsive for same sublattice, attractive for different sublattices \cite{aj2012}. This mechanism for repulsive Casimir force can be contrasted with the system described in \cite{grushin} wherein a topological insulator is used as a dielectric medium, and in which the sign of the interaction is varied by tuning the magnetoelectric polarizability of the medium.
The Casimir effect is strongest for the horizontal configuration, where the domain walls are rows of closely spaced same-sublattice flips. All other orientations of the domain walls lead to faster decay. In particular, a domain wall along the $\vec{r}_1+\vec{r}_2$ direction for example (Fig.\label{twolines.fig}), which alternates flips on the A- and B-sublattices, leads to a short range contact-type interaction.

\begin{figure}[t]
\begin{center}
\includegraphics[scale=0.7]{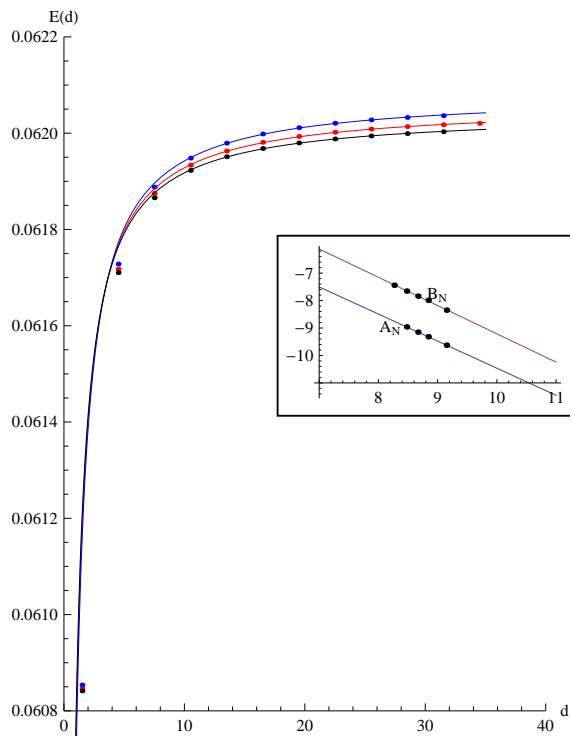}
\caption{
Change in ground state energy per unit length $e(d)$ plotted as a function of the distance $d$ between two rows of defects for three system sizes ($N=4800,5808,6912$). The lines represent a fit to Eq.\ref{casim.eq}. The inset shows in a log-log plot the scaling with $1/N$ of the $A_N$ and $B_N$ coefficients.}
\label{casim.fig}
\end{center}
\end{figure}

In sum, the interaction energy $E(d)$ between horizontal domain walls depends on the slab length $L_x$ and thickness $d$ as $-L_x/d$. The proportionality to slab length is expected whereas the dependence on $d$ is not obvious. The inverse-$d$ dependence of $E(d)$ can be deduced from a dimensional analysis as follows. On the one hand, the interaction energy is a dimensionless quantity when expressed in units of $J$ (the only energy scale in the problem), and on the other hand, this energy is expected to be proportional to the length of the interface, which the interaction energy is proportional to $L_x/d$. This distance dependence of the energy can be compared with that of the electromagnetic field in a similar geometry. In that case, a dimensional analysis predicts that the interaction energy due to quantum fluctuations, which is proportional to $\hbar c$, should drop off as $L_x/d^2$. Our model also differs in several essential respects from the one considered in \cite{hone} between two rows of holes in a square lattice antiferromagnet, where a Casimir effect is observed. For that problem, the interaction, always attractive was found to decay as $1/d^2$. In that case, each row of holes is characterized by the microscopic length scale $a$, and the region between is the unperturbed antiferromagnet. In contrast, the slab geometry that we consider separates two regions which are distinct, although related by phason flips.

A back-of-the-envelope calculation for this inverse distance dependence of $E$ can also be given. We note that at long distance, the two phason interaction energy decays as $r^{-3}$ \cite{aj2012}. The $d$ dependence of the energy of the slab of phasons can be estimated by computing the energy of interaction of a single A-sublattice phason on the edge due to its interactions with successive rows of B-sublattice phasons, which is proportional to $\int_0^d dy\int_0^{L_x/2} dx (x^2+y^2)^{-3/2} \sim 1/d$ for $L_x \gg d \gg a$. To this one should add a second energy contribution of the opposite sign, due to the interactions of $same$ sublattice phasons, but the distances are different and therefore the contributions will not completely cancel. This rough calculation for the energy per unit length makes use of the simplified asymptotic form of the 2-phason interaction, and does not take into account the detailed angular dependence of the anisotropic decay.

We address finally the question of possible experimental realizations of the ordinary dice and staggered dice lattices. The ordinary dice lattice can been constructed using optical trapping of cold atoms \cite{berc}, thus providing an experimental realization of electron tight binding models with so-called Dirac-Weyl fermions. Cold trapped atoms could similarly also be used to realize the Heisenberg model on the dice lattice, and possibly even the staggered dice lattice which is a more complex structure. In conventional condensed matter systems, the dice lattices could perhaps be realized in layered compounds in a way analogous to the realization of the Heisenberg model on the two dimensional Kagome lattice (dual of the dice lattice). In the mineral named volborthite, for example, the spins are associated with Cu atoms, interacting via super-exchange processes. When each of these magnetically active layers is viewed from above, the spins and the bonds are seen to project to the Heisenberg antiferromagnetic model on a Kagome lattice \cite{volbor}. The so-called checkerboard model can be realized by a mapping in which projecting spins on a three dimensional pyrochlore lattice \cite{starykh}.  In an analogous way, our model on the dice lattice
could be realized by considering spins on vertices of cubes, and considering their projection on a plane perpendicular to one of the body diagonals that bisects the cube. The six sites of the hexagonal cell of the dice lattice correspond to the six vertices lying closest to this plane (shown in red in Fig.\ref{proj.fig} for a row of cubes). The two remaining apical vertices shown in blue, both project onto the central site in the cage but with different connectivities. If the one of these configurations is selected randomly, for example, one obtains the phason-disordered case. To realize the parallel plate geometry discussed in this paper, one could in principle suitably engineer the substrate on which the spin system is built, so as to slightly bias one or other of the two configurations. Transitions could then be induced by small changes of the distances of the apical atoms. Yet another approach could consist of using molecular magnets as building blocks to construct the dice lattice systems, using the methods described in \cite{molmag}.

\begin{figure}[t]
\begin{center}
\includegraphics[scale=0.5]{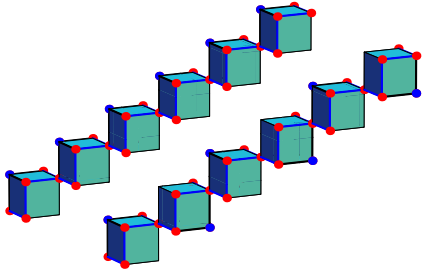}
\includegraphics[scale=0.5]{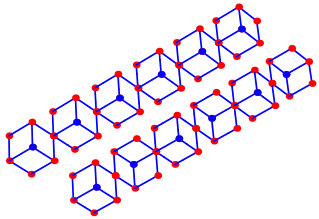}
\caption{(upper) Two examples of rows of cubes showing the vertices before projection. Red sites project onto hexagonal cage vertices, while the blue site can lie either above or below the 3-fold plane bisecting the cubes. The upper row shows a uniform choice of blue sites, and the lower row a staggered choice. (lower)The sites after projection on the three-fold plane.  The decoration is uniform in all cells in the upper row, while in the lower row it is staggered.
 }
\label{proj.fig}
\end{center}
\end{figure}

In conclusion, we have computed the Casimir forces in a two dimensional antiferromagnet based on the honeycomb lattice. The fluctuations have a geometrical origin, involving changes of local topology, and may be realized in quasi-two-dimensional structures in which the nearest neighbor environments allow for local transitions between two different conformations with little or no energy cost. The interaction energy depends on the orientation of the Casimir slab, and can be attractive or repulsive depending on whether the walls belong to the same to different sublattices. The effect is strongest for a slab with horizontal orientation, the interaction energy falling off as the inverse of the slab thickness.  The magnitude of the effect is estimated to be of the order of $pJ/m$, for  antiferromagnetic couplings $J$ of the order of 1$eV$. This fluctuation-induced interaction could be measured for sheets of spins in which the phason flips are flips between isomeric configurations.

\begin{acknowledgments} We would like to thank B. Dou\c{c}ot and Thorsten Emig for useful discussions.
\end{acknowledgments}


\begin{references}
\bibitem{casimir} H.B.G. Casimir, Proc. Kon. Ned. Akad. Wetensch. B51, 793 (1948), M. Kardar and R. Golestanian, Rev. Mod. Phys. {\bf 71}, 1233, (1999).

\bibitem{kardar2} M. Kardar and R. Golestanian, Rev. Mod. Phys. {\bf 71} 1233 (1999)

\bibitem{kardar1} H. Li and M. Kardar, P.R.L. {\bf 67} 3275 (1991); H. Li and M. Kardar, P.R.A {\bf 46} 6490 (1992)

\bibitem{williams} G.A. Williams, P.R.L. {\bf 92} 197003 (2004)

\bibitem{ajdari} A. Ajdari, P.R.L. {\bf 66} 1481 (1991)

\bibitem{cattuto} C. Cattuto et al, P.R.L. {\bf 96} 178001 (2006)


\bibitem{tls} W. Phillips, J. Low Temp. Phys. {\bf 7} 351 (1972); P.W. Anderson, B.I. Halperin and C.M. Varma, Phil. Mag. {\bf 25} 1 (1972)

\bibitem{note} The term ``phason flip" refers to a single local change of the lattice in this paper, while the term ``phason", coined as an analogy with ``phonon", usually denotes a collective excitation in quasicrystals (M. Widom, Philosophical Magazine {\bf 88} 2339 (2008)).

\bibitem{aj2012} A. Jagannathan, B. Dou\c{c}ot, A. Szallas and S. Wessel, Phys. Rev. B {\bf 85}, 094434, (2012).


\bibitem{kato} K. Kato, S. Todo, K. Harada, N. Kawashima, S. Miyashita and H. Takayama, Phys. Rev. Lett. {\bf 84} 4204 (2000).

\bibitem{sand} A. W. Sandvik, Phys. Rev. B {\bf 66}, 024418 (2002).

\bibitem{lin} Y.C. Lin, R. Melin, H. Rieger and F. Igloi, { Phys. Rev.} B {\bf 68} 024424 (2003).

\bibitem{flor} N. Laflorencie, S. Wessel, A. L\"auchli, and H. Rieger,{ Phys. Rev.} B {\bf 73} 060403(R) (2006).

\bibitem{mucc} E.R. Mucciolo, A.R. Castro Neto and C. Chamon, Phys. Rev. B {\bf 69} 214424 (2004).

\bibitem{footnote} The term "frustrated" is used when it is not possible to simultaneously minimize the energies of all pairs of nearest neighbor spins, as occurs, for example, for three spins on vertices of a triangle.

\bibitem{villain} J. Villain, R. Bidaux, J-P. Carton and R. Conte, J. Physique {\bf 41} 1263 (1980)

\bibitem{prb2007}A.~Jagannathan, R.~Moessner  and S.~Wessel, Phys. Rev. B, {\bf 74} 184410, (2006).

\bibitem{hols} T. Holstein and H. Primakoff, Phys. Rev. {\bf 58}, 1098 (1940).


\bibitem{bulut} N. Bulut, D. Hone, D.J. Scalapino and E.Y. Loh, Phys. Rev. Lett. {\bf 62}, 2192, (1989).

\bibitem{grushin} Adolfo G. Grushin and Alberto Cortijo, Phys. Rev. Lett. {\bf 106}, 020403, (2011).

\bibitem{hone} L.P. Pryadko, S. Kivelson and D. W. Hone, Phys. Rev. Lett. 80, 5651–5654 (1998) ;  D. W. Hone, S. Kivelson and L.P. Pryadko, Stripes and related phenomena, Selected topics in superconductivity, vol.8, 447 (Springer 2002)

\bibitem{berc} D. Bercioux, D.F Urban, H. Grabert and W. H\"ausler, {Phys. Rev.} A {\bf 80} 063603 (2009).

\bibitem{volbor} Z. Hiroi, J. Phys. Soc. Japan {\bf 70} 3377 (2001)

\bibitem{starykh} O.A. Starykh, A. Furusaki and L. Balents, Phys. Rev. B  72 , 094416 (2005).

\bibitem{molmag} Ie-Rang Jeon and   Rodolphe Clérac, Dalton Trans., 41, 9569 (2012)

\end{references}
\end{document}